\documentclass[prl,letter,twocolumn,showpacs,superscriptaddress,floatfix]{revtex4-1}
\usepackage{hyperref}
\usepackage{amssymb}
\usepackage{amsmath}
\usepackage{float}
\usepackage{bbm}
\usepackage{graphicx}
\usepackage{epstopdf}
\usepackage{epsfig}
\usepackage[usenames]{color}
\usepackage{bbold}

\pacs{02.70.Ss, 
02.70.Tt,
05.10.Ln
}
\begin{document}
\newcommand{\tkmax}{t_K^\text{max}}
\newcommand{\Gup}{\ensuremath{G_{|\uparrow\rangle}}}
\newcommand{\Gdn}{\ensuremath{G_{|\uparrow\rangle}}}
\newcommand{\Gzero}{\ensuremath{G_{|0\rangle}}}
\newcommand{\Gupdn}{\ensuremath{G_{|\uparrow\downarrow\rangle}}}
\newcommand{\Gj}{\ensuremath{G_{|j\rangle}}}
\newcommand{\Gzeroj}{\ensuremath{G^0_{|j\rangle}}}
\newcommand{\Deltalesser}{\ensuremath{\Delta^<}}
\newcommand{\Deltagreater}{\ensuremath{\Delta^>}}
\newcommand{\Sigmazero}{\ensuremath{\Sigma_{|0\rangle}}}
\newcommand{\Sigmaup}{\ensuremath{\Sigma_{|\uparrow\rangle}}}
\newcommand{\Sigmadn}{\ensuremath{\Sigma_{|\downarrow\rangle}}}
\newcommand{\Sigmaupdn}{\ensuremath{\Sigma_{|\uparrow\downarrow\rangle}}}
\newcommand{\Sigmaj}{\ensuremath{\Sigma_{|j\rangle}}}
\newcommand{\GSigmaj}{\ensuremath{[\Gj\Sigmaj]}}
\newcommand{\SigmaGj}{\ensuremath{[\Sigmaj\Gj]}}
\newcommand{\Fijs}{\ensuremath{F^\sigma_{|i\rangle\langle j|}}}
\newcommand{\Fijup}{\ensuremath{F^\uparrow_{|i\rangle\langle j|}}}
\newcommand{\Fijdn}{\ensuremath{F^\downarrow_{|i\rangle\langle j|}}}
\newcommand{\Fijsdagger}{\ensuremath{F^{\sigma\dagger}_{|i\rangle\langle j|}}}
\newcommand{\Fsdagger}[2]{\ensuremath{F^{\sigma\dagger}_{|#1\rangle\langle #2|}}}
\newcommand{\Fs}[2]{\ensuremath{F^\sigma_{|#1\rangle\langle #2|}}}
\newcommand{\G}[1]{\ensuremath{G_{|#1\rangle}}}
\newcommand{\K}[4]{\ensuremath{K_{|#1\rangle|#2\rangle}(#3,t_F,#4)}}
\newcommand{\J}[4]{\ensuremath{J_{|#1\rangle|#2\rangle}(0,#3;#4,\tkmax)}}

\author{Emanuel Gull}
\affiliation{Department of Physics, Columbia University, New York, New York 10027, USA}
\author{David R. Reichman}
\affiliation{Department of Chemistry, Columbia University, New York, New York 10027, USA}
\author{Andrew J. Millis}
\affiliation{Department of Physics, Columbia University, New York, New York 10027, USA}

\title{Numerically Exact  Long Time  Behavior of Nonequilibrium Quantum Impurity Models}

\date{\today}

\hyphenation{}

\begin{abstract}
A Monte Carlo sampling of diagrammatic corrections to the non-crossing approximation is shown to provide numerically exact estimates of the  long-time dynamics and steady state properties of nonequilibrium quantum impurity models. This `bold' expansion converges uniformly in time and significantly ameliorates the sign problem  that has heretofore limited the power of real-time Monte Carlo approaches to strongly interacting real-time quantum problems.  The new approach enables the study of previously intractable problems ranging from generic long time nonequilibrium transport characteristics in systems with large onsite repulsion to the direct description of spectral functions on the real frequency axis in Dynamical Mean Field Theory.
\end{abstract}

\maketitle

Numerical evaluation of diagrammatic perturbation series has played an important role in fields including quantum electrodynamics,\cite{Kinoshita90}  statistical mechanics \cite{Sandvik10} and condensed matter physics.\cite{Prokofev98,Kozik10,Rubtsov05,Werner06,Gull08_ctaux,GullRMP}  Recent developments have established diagrammatic Monte Carlo methods \cite{GullRMP}  as particularly  powerful tools for the study of finite size clusters of interacting fermions coupled to a noninteracting bath. These `quantum impurity models' are  used to model the physics of nanosystems coupled to leads \cite{Hanson07}, adsorption of atoms on surfaces \cite{Brako81} and  magnetic impurities in metals,\cite{Anderson61} and are important as auxiliary problems in dynamical mean field theory approaches to infinite lattice correlated systems.\cite{Georges96,Maier05}

Diagrammatic Monte Carlo methods have been very successful for equilibrium problems, where the Hamiltonian $H$ is partitioned as $H=H_0+H_1$ the partition function $Z$ at temperature  $T$ is expressed in an interaction representation as $Z=\text{Tr}T_\tau \exp\left[-\int_0^{1/T} H_1(\tau)\right]$ and the exponential is expanded. The diagrammatic expansion order needed to obtain reliable results grows as $\langle H_1\rangle/T$, setting  a lower limit on the temperatures which can be studied with fixed computational resources, but in practice the method works down to  temperatures which are very low relative to the basic scales \cite{Gull07}. The use of Wick's theorem,  which organizes fermion contractions into determinants, dramatically reducing the number of diagrams which must be sampled and eliminating one source of fermion sign problem, is crucial to the success of the procedure.

In nonequilibrium problems the role of the partition function is played by the time evolution operator $K\sim \exp\left[iH_1t\right]$  and the role of inverse temperature is played by the time interval $t$ to be studied. The factor of $i$ means that a straightforward expansion suffers from a  phase problem, which in practice severely limits the diagrammatic order which can be sampled and therefore the time intervals which can be studied. To date only relatively short times (up to $\sim 2-3$ times the hybridization scale) have been accessed \cite{Muehlbacher08,Schiro09,Werner09,Werner10}. 

In analytic many-body theory, partial resummation techniques are often used to sum up specific classes of diagrams. If the resummation captures enough of the physics, one may hope that an expansion around it will converge rapidly. Motivatived by this idea, we formulated such an expansion for equilibrium properties of quantum impurity models \cite{Gull10_bold}.  We found that while the method lead to a marked decrease in the mean perturbation order, the gain was in most cases offset by a sign problem, which is generic to this class of methods and arises from the absence of Wick's theorem for expansions about partially resummed theories.

In this paper we present a new formulation of a numerical expansion around a partial resummation, applicable to nonequilibrium quantum impurity models and therefore also to time dependent dynamical mean field theory. We refer to this expansion as a `bold' expansion but note that it differs from algorithms (also termed `bold expansions') which sample higher-order self-energy diagrams based on numerically computed lower-order diagrams but employ no partial resummations \cite{Prokofev07,Prokofev08B}. Our method is based on a stochastic sampling of the full configuration space and is numerically exact. New aspects of our work include a treatment of the vertex corrections essential for the evaluation of expectation values, and most importantly a demonstration that in contrast to the bare expansion, the convergence of the bold expansion is {\em uniform in time}: even the long time behavior is adequately characterized by {\em finite} orders of bold perturbation theory. We present results for the steady state density matrix, charge and magnetic relaxation times  and  current of a model quantum dot which demonstrate that the method allows access to unprecedentedly long times.  

We demonstrate the power of the method on the  Anderson model with Hamiltonian
\begin{align}
H_A=\sum_\sigma\left(\varepsilon_d+H\sigma\right)d^\dagger_\sigma d_\sigma+Un_\uparrow n_\downarrow+H_\text{hyb}+H_\text{lead}\label{H}
\end{align}
which describes a quantum dot with a  single spin-degenerate orbital with correlation energy $U$ hybridized  to two leads labeled by $a=L,R$. The Hilbert space of the  impurity consists of four states: $|0\rangle, |\uparrow\rangle,|\downarrow\rangle$ and $|\uparrow\downarrow\rangle$. $H$ describes a magnetic field directed parallel to the spin quantization axis,
$H_\text{hyb}=\sum_{ka\sigma}\left[V_{ka\sigma}d^\dagger_\sigma c_{ka\sigma}+H.c.\right]$ parametrizes the hybridization between the level and the leads  and $H_\text{lead}$ describes the dynamics of the  leads.  Lead $a$ is assumed to be in equilibrium at chemical potential $\mu_a$ and temperature $T_a$; the presence of two leads allows for departures from equilibrium parametrized by  $\mu_L\neq\mu_R$ or $T_L\neq T_R$. An important parameter is the level width $\Gamma=\sum_{ka}  V_{ka}^2\delta(\varepsilon_{ka}-\mu_a)$. For our specific calculations we use the parametrization of Ref.~\cite{Werner10} with  $\mu_L=-\mu_R = V/2, $ $\nu=10=\omega_c.$

We wish to compute time dependent expectation values of  operators ${\hat O}$ such as the dot charge $(n)$ and spin $(m)$ densities $n_\uparrow\pm n_\downarrow$ or the current flowing into the dot from (say) the left lead $J_L=i\sum_{k\sigma}\left[V_{kL\sigma}d^\dagger_\sigma c_{kL\sigma}-H.c.\right]$. These may be obtained from  the time dependent density matrix  ${\hat \rho}(t)$ as:
\begin{align}
\langle{\hat O}(t_F)\rangle=\text{Tr}\left[{\hat O}{\hat \rho}(t_F)\right]=\text{Tr}\left[{\hat O}e^{-iH_At_F}{\hat \rho}_0e^{iH_At_F}\right].
\label{expectation}
\end{align}  
For non-equilibrium problems the only approach  known to be reliable is to compute ${\hat \rho}(t)$ by evolving forward from an initial condition ${\hat \rho}_0$ as in the second term of Eq.~\ref{expectation}.

\begin{figure}[t]
\begin{tabular}{lclc}
a)&{\includegraphics[width=0.45\columnwidth]{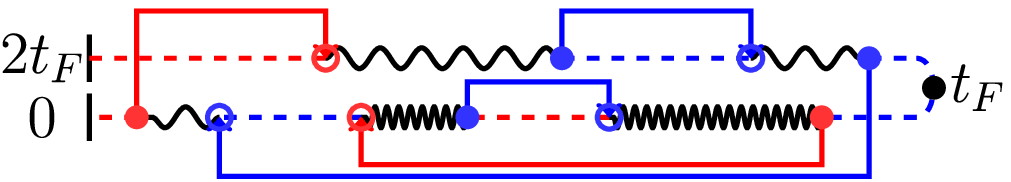}}&
b)&{\includegraphics[width=0.45\columnwidth]{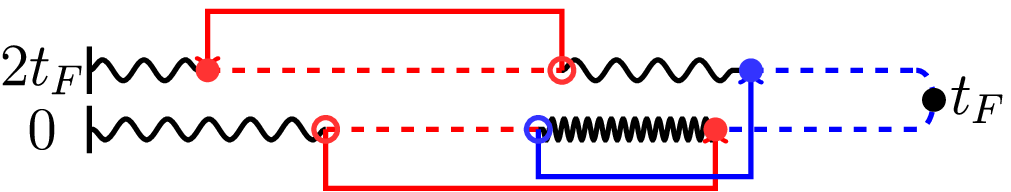}}\\
c)&{\includegraphics[width=0.45\columnwidth]{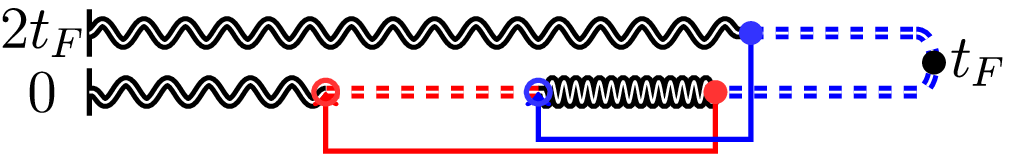}}&
d)&{\includegraphics[width=0.45\columnwidth]{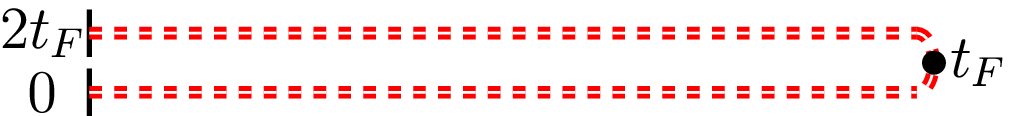}}\\
e)&{\includegraphics[width=0.45\columnwidth]{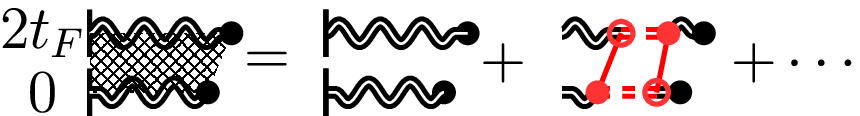}}&
f)&{\includegraphics[width=0.45\columnwidth]{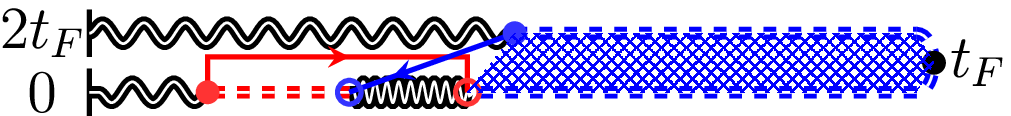}}
\end{tabular}
\caption{Diagrams arising in the bare hybridization and `bold' expansions  of Eq.~\ref{H} on the Keldysh contour: propagation in eigenstates of $H_\text{dot}$ is described by wavy  and dashed lines. Solid lines denote hybridization functions, and circles hybridization vertices. $a)$ NCA diagram. $b)$ diagram containing crossing. $d)$ bold propagator (dashed double line) which resums $a)$. $c)$ Diagram of bold expansion which resums diagrams including $b)$.
Diagram $e)$ illustrates vertex resummations, and diagram $f)$ shows a bold vertex diagram.
}
\label{diagrams2}
\end{figure}

We take  ${\hat \rho}_0 = {\hat \rho}_0^\text{dot}\otimes {\hat \rho}^\text{lead}$ corresponding to decoupled impurity and leads and assume that  ${\hat\rho}_0^\text{dot}$ is diagonal in the occupation number basis. We evaluate Eq.~\ref{expectation} by writing the time evolution operators $e^{\pm itH_A}$ in an interaction representation with respect to $H_\text{hyb}$ and  expanding powers of $H_\text{hyb}$. The bare expansion produces  diagrams of the form shown in panels (a) and (b)  of Fig.~\ref{diagrams2}. The presence of two time evolution operators in Eq.~\ref{expectation} means that two time contours are required, one running from an initial time $t=0$ (left side of lower contour) to the measurement time $t_F$ (right side) and the other running back to initial time (left side of upper contour, label $2t_F$ indicating total time interval along double contour). Hybridization vertices $V_{ka\sigma}d^\dagger_\sigma c_{ka\sigma}$ $(V_{ka\sigma}^{*} c^\dagger_{ka\sigma}d_\sigma)$ occurring at times $t_1...t_j$ are indicated by heavy (empty) dots as in Ref.~\cite{Gull10_bold} and are connected by light lines displaced from the basic contour indicating contractions of the lead ($c$) operators computed using ${\hat \rho}^\text{lead}$ and by solid, wavy or dashed lines indicating propagation in  eigenstates of $H_\text{dot}$.

A straightforward evaluation of Eq.~\ref{expectation} thus requires a sum over all diagrams, a sum over all contractions of lead operators, and an integral over all times. The  sign problem arising from the powers of $i$ limits diagrammatic Monte Carlo studies  to situations where the mean perturbation order is $\lesssim 10$  \cite{Muehlbacher08,Schiro09,Werner09}    and for this reason only times $t\lesssim 2/\Gamma$ are accessible with this method. We therefore investigate a bold method.  As an analytic resummation we use the non-crossing approximation (NCA) \cite{Keiter71,Bickers87} but we emphasize that the concepts and methods developed here apply to any bold expansion.

The NCA integrates diagrams without `crossing' hybridization lines [Fig.~\ref{diagrams2}$a$ shows a sample] using coupled integral equations. The correction diagrams are obtained along lines similar to those of the equilibrium algorithm of Ref.~\cite{Gull10_bold}. Bare `atomic state' propagators and non-crossing hybridization lines are replaced by `bold' NCA propagators (denoted by heavy lines in Fig.~\ref{diagrams2}). Diagrams with non-crossing hybridization lines contained in the underlying NCA propagators are not sampled. Thus for example the bold sampling process collapses diagram (a) of Fig.~\ref{diagrams2} to diagram (d), and diagram (b) to diagram (c). 

In defining crossing and non-crossing diagrams the issue of lines connecting one contour to another arises. These lines may be interpreted as vertex corrections to the operator placed at the measurement point $t_F$ and to the initial density matrix.  While they can be sampled directly, we find that it is best to resum non-crossing lines spanning a contour into NCA vertex corrections, and to replace the bare operator by the combination of the operator and its NCA vertex correction. An example of a vertex correction is shown in panel (e) of Fig. \ref{diagrams2} and  the resulting `bold' version  in panel (f). Vertex corrections, especially for the vertices spanning the initial density matrix, significantly reduce the expansion order and the dynamic sign problem and allow us to perform an expansion about the NCA steady state.

Our Monte Carlo process is defined by moves which propose the addition or removal of vertices on either contour. The proposals are made without regard to whether the diagram is bold or not, but a proposed move which produces a diagram which is subsumable into a bold diagram is rejected.  The procedure is exact because each bare diagram is contained in exactly one bold diagram.  We combine diagrams in such a way that all terms are real \cite{Werner09} so the phase problem becomes a sign problem. For computations of a given observable ${\hat O}$ the acceptance/rejection probabilities of a given move are determined from the absolute value of the contribution to $\langle{\hat O}\rangle$ and one measures $\langle{\hat O}\ \text{sign}\rangle/\langle \text{sign}\rangle$.   The expectation value of the sign decays exponentially with perturbation order considered and thus with the time interval to be studied. Management of the sign problem is a crucial issue in this and related methods. 

We have found it  useful to define a  diagrammatic configuration at expansion order $k$  (i.e. a set of vertices at times $t_1...t_{2k}$) as the sum of all  contractions of lead  operators consistent with the  crossing condition. The lack of a Wick's theorem means that the  sum  must be performed explicitly. The exponential growth with perturbation order of the number of possible  contractions sets a limit $\sim 10$ on the order which can be reached, but this limit is less severe in practice than the limit imposed by the sign. It is possible that higher orders may be reached by integrating diagrams individually or combining only a subset of them; this has not yet been explored.

\begin{figure}[t]
\includegraphics[width=0.9\columnwidth]{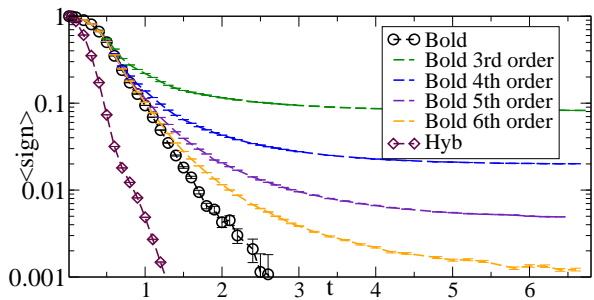}\\
\caption{Average sign as a function of time in evaluation of expectation value of  current for $U=4,\beta=50,H=0,V=5.$ Bare expansion: diamonds. Bold expansion: circles. Other lines: Bold expansion truncated at orders $3$ to $6$. }
\label{sign_problem}
\end{figure}

Fig.~\ref{sign_problem} shows the time dependence of the expectation value of the sign computed in an expansion of the current for  the nonequilibrium Anderson model. The diamonds show the sign obtained from the bare hybridization expansion method  of Refs  \cite{Muehlbacher08,Schiro09,Werner09} and the circles the sign obtained from a straightforward application of the bold method.  (Essentially identical sign vs time curves are found for all parameters studied except that $\langle \text{sign}\rangle$ increases at very high $T \gtrsim \Gamma$.) The larger mean value of the sign at a given time in the bold method arises because fewer perturbation orders are needed to reach a solution. The exponential decrease of $\langle \text{sign} \rangle$ with time visible in Fig.~\ref{sign_problem}  constrains the times that can be studied with finite resources. We see that the straightforward bold expansion can reach $\approx$ twice as long a time as the bare expansion.

\begin{figure}[tb]
\includegraphics[width=0.9\columnwidth]{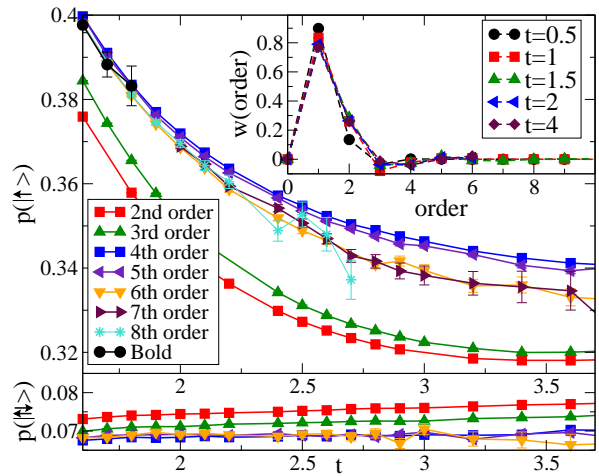}
\caption{Order by order convergence for times $t=1.5,...,4$. 
Upper panel: spin imbalanced parameters, density expansion, $U=4, V=5$, state $|\uparrow\rangle$ (initial state $|\uparrow\rangle$), 
blowup to region of biggest differences. Lower panel: typical case, $U=8, V=2, H=0$, state $|\uparrow\downarrow\rangle$ starting from $|\uparrow\rangle$. 
Inset: order-by-order contribution to current, for  $U=4, V=5$ and times indicated.
}\label{order_by_order_states}
\end{figure}

In contrast to equilibrium  simulations, where the diagrams generated by the Monte Carlo process are typically the ones most important to the evaluation of the observable, we find that in the nonequilibrium situations considered here an unconstrained Monte Carlo exploration of bold diagrams generates many high order diagrams which sum to zero in the observable. Thus we define a Monte Carlo process which considers only diagrams with perturbation order less than or equal to some value $k$, and then  increase $k$ until convergence is reached.  For the cases studied here a $k\lesssim 8$ sufficed.  Fig.~\ref{sign_problem} shows that the mean sign decreases exponentially with increasing maximum perturbation order but for a given perturbation order saturates at a non-zero value. We find that once the correct $k$ is identified, the bold expansion can be arranged so that convergence is {\em uniform in time:} the mean perturbation order required to obtain a convergent result does not increase as the time interval is increased. 

Convergence is poorest for spin-dependent properties of spin-imbalanced initial conditions; the main panel of Fig.~\ref{order_by_order_states} shows the slowest-converging case we have encountered so far. The straightforward bold expansion only converges out to times $t\approx 2$, the convergence with order is oscillatory but by $7^{th}$ order an acceptable convergence is reached as can be seen from the coincidence of the $6^{th}, 7^{th}$, and $8^{th}$ order results. The lower panel shows a more typical case, where convergence is monotonic and occurs by $4^{th}$ order. The upper inset presents the contribution $w(t)$ made to the current at time $t$ by the sum of all diagrams of a given order. We see that for this case diagrams of order $\gtrsim 5$ make no net contribution, but would be extensively sampled in a straightforward bold Monte Carlo calculation. 

\begin{figure}[tb]
\includegraphics[width=0.9\columnwidth]{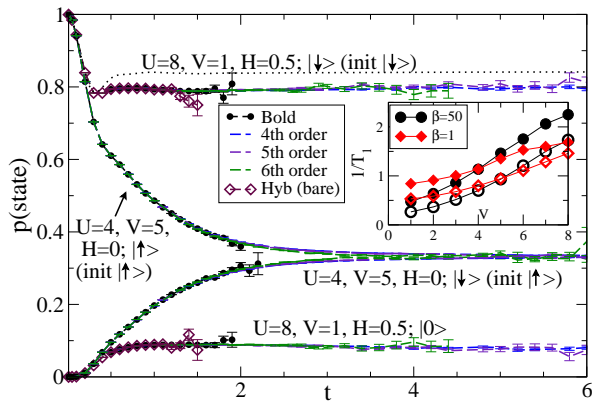}
\caption{
Time evolution of dot states from specified initial condition, for parameters indicated calculated using bare expansion, bold expansion, and truncated bold expansion. Dotted line: NCA.
Inset: Decay rate $1/T_1$ calculated for nonequilibrium Anderson model at voltages and temperatures indicated for $U=8.$ Open symbols: NCA. Filled symbols: Bold.
}
\label{state0}
\end{figure}

\begin{figure}[tb]
\includegraphics[width=0.9\columnwidth]{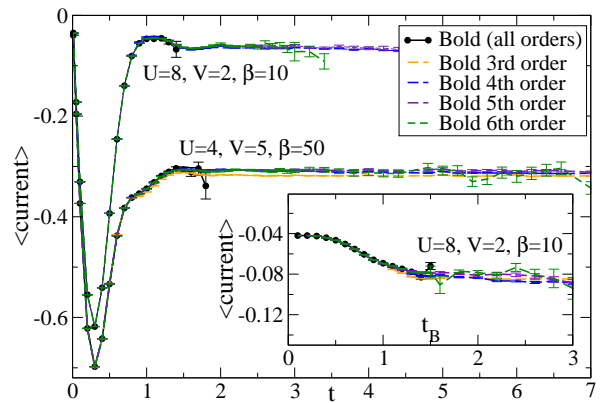}
\caption{
Time evolution of current for parameters indicated. Main panel: starting from empty dot. Inset: starting from the NCA steady state.
}
\label{decayU4}
\end{figure}

The much longer times  accessible via the methods proposed here allow us to reach physically interesting steady states. Fig.~\ref{state0} shows the evolution of particular diagonal elements of the density matrix for different model parameters and starting from different initial conditions.  The top and bottom traces $(U=8, V=1, H=0.5$) show the evolution of the spin down (favored by $H$) and empty states starting from the initial condition in which the dot is in $|\downarrow\rangle.$ The similar time scales in the evolution of the empty and singly occupied states show that the time dependence, which is rapid and is captured correctly by the bare and straightforward bold methods, is almost entirely due to charge relaxation.  By contrast, the middle traces $(U=4, V=5, H=0)$ show the evolution of spin up and down states from a spin polarized initial condition. The much slower spin relaxation is evident. The times $t\gtrsim 3$ required to access the steady state are only accessible by the new methods proposed here. Similarly, Fig.~\ref{decayU4} shows the time evolution of the current.

For a wide range of parameters and initial conditions we find that the magnetization relaxes exponentially to its steady state value:  $m\sim \exp[-t/T_1]$. The inset of Fig.~\ref{state0} compares the voltage and temperature dependence of the spin relaxation rate computed in the bold expansion and the NCA. The latter systematically underestimates relaxation rates. Remarkably, the temperature dependence of $T_1$ is opposite at high and low voltages.

The presence of analytically computed vertex functions in the algorithm allows improved access to the steady state by starting the bold computation from the density matrix corresponding to the NCA steady state rather than from   a decoupled or non-interacting initial condition.
In practice we use the NCA equations to propagate forward for a time $t_0$ from a decoupled state, after which the bold interactions are turned on and a further time $t_B$ is studied. 
For the parameters we have studied the NCA density matrix is typically close to the true steady state with the largest differences occurring for a non-zero field (dotted line, Fig.~\ref{state0}). Transients decay quickly.
While NCA propagators and vertices are  required for the entire time interval $t_F=t_0+t_B$ the bold expansion need only operate over the much shorter time $t_B$. The inset of Fig.~\ref{decayU4} shows the time evolution of the current from the NCA steady state to the numerically exact steady state for a representative choice of parameters. The large initial transient observed in the main panel is absent. 

In conclusion, we have developed a real time diagrammatic method that enables a description of long time and  steady-state properties in nontrivial quantum impurity models such as the equilibrium and nonequilibrium Anderson models over a wide range of interaction strengths and time scales. The approach is based on a systematic summation of terms contained in an expansion in powers of the hybridization portion of the Hamiltonian about a state described by an analytical resummation. This `bold expansion' is numerically exact, uniformly convergent and greatly reduces the real time sign problem that inhibits the study of long time properties in `bare' continuous time quantum Monte Carlo methods. We found that in many cases the non-crossing approximation provides reasonably accurate ($\lesssim 5\%$) estimates of the diagonal elements of the steady-state density matrix, but is less reliable for relaxation rates.  Future work will be devoted to using the method described here as a real-time impurity solver for DMFT (where two-time correlators are required) and applying it to a range of physically relevant situations. A crucial question is how far into the Kondo regime the method can be pushed. 

\acknowledgments
We  thank P.~Werner for many discussions and helpful insights. EG and AJM are supported by NSF-DMR-1006282, DRR by NSF-CHE-0719089. Calculations were performed on Brutus at ETH Zurich, using a code based on the ALPS\cite{ALPS2} library. A portion of this research was conducted at the Center for Nanophase Materials Sciences, which is sponsored at Oak Ridge National Laboratory by the Office of Basic Energy Sciences, U.S. Department of Energy.
\bibliography{refs_shortened.bib}

\begin{thebibliography}{24}%
\makeatletter
\providecommand \@ifxundefined [1]{%
 \@ifx{#1\undefined}
}%
\providecommand \@ifnum [1]{%
 \ifnum #1\expandafter \@firstoftwo
 \else \expandafter \@secondoftwo
 \fi
}%
\providecommand \@ifx [1]{%
 \ifx #1\expandafter \@firstoftwo
 \else \expandafter \@secondoftwo
 \fi
}%
\providecommand \natexlab [1]{#1}%
\providecommand \enquote  [1]{``#1''}%
\providecommand \bibnamefont  [1]{#1}%
\providecommand \bibfnamefont [1]{#1}%
\providecommand \citenamefont [1]{#1}%
\providecommand \href@noop [0]{\@secondoftwo}%
\providecommand \href [0]{\begingroup \@sanitize@url \@href}%
\providecommand \@href[1]{\@@startlink{#1}\@@href}%
\providecommand \@@href[1]{\endgroup#1\@@endlink}%
\providecommand \@sanitize@url [0]{\catcode `\\12\catcode `\$12\catcode
  `\&12\catcode `\#12\catcode `\^12\catcode `\_12\catcode `\%12\relax}%
\providecommand \@@startlink[1]{}%
\providecommand \@@endlink[0]{}%
\providecommand \url  [0]{\begingroup\@sanitize@url \@url }%
\providecommand \@url [1]{\endgroup\@href {#1}{\urlprefix }}%
\providecommand \urlprefix  [0]{URL }%
\providecommand \Eprint [0]{\href }%
\providecommand \doibase [0]{http://dx.doi.org/}%
\providecommand \selectlanguage [0]{\@gobble}%
\providecommand \bibinfo  [0]{\@secondoftwo}%
\providecommand \bibfield  [0]{\@secondoftwo}%
\providecommand \translation [1]{[#1]}%
\providecommand \BibitemOpen [0]{}%
\providecommand \bibitemStop [0]{}%
\providecommand \bibitemNoStop [0]{.\EOS\space}%
\providecommand \EOS [0]{\spacefactor3000\relax}%
\providecommand \BibitemShut  [1]{\csname bibitem#1\endcsname}%
\let\auto@bib@innerbib\@empty
\bibitem [{\citenamefont {Kinoshita}(1990)}]{Kinoshita90}%
  \BibitemOpen
  \bibinfo {editor} {\bibfnamefont {T.}~\bibnamefont {Kinoshita}},\ ed.,\
  \href@noop {} {\emph {\bibinfo {title} {Quantum Electrodynamics}}}\ (\bibinfo
   {publisher} {World Scientific},\ \bibinfo {year} {1990})\BibitemShut
  {NoStop}%
\bibitem [{\citenamefont {Sandvik}(2010)}]{Sandvik10}%
  \BibitemOpen
  \bibfield  {author} {\bibinfo {author} {\bibfnamefont {A.~W.}\ \bibnamefont
  {Sandvik}},\ }\href {\doibase 10.1063/1.3518900} {\bibfield  {journal}
  {\bibinfo  {journal} {AIP Conf. Proc.}\ }\textbf {\bibinfo {volume} {1297}},\
  \bibinfo {pages} {135} (\bibinfo {year} {2010})}\BibitemShut {NoStop}%
\bibitem [{\citenamefont {Prokof'ev}\ and\ \citenamefont
  {Svistunov}(1998)}]{Prokofev98}%
  \BibitemOpen
  \bibfield  {author} {\bibinfo {author} {\bibfnamefont {N.~V.}\ \bibnamefont
  {Prokof'ev}}\ and\ \bibinfo {author} {\bibfnamefont {B.~V.}\ \bibnamefont
  {Svistunov}},\ }\href {\doibase 10.1103/PhysRevLett.81.2514} {\bibfield
  {journal} {\bibinfo  {journal} {Phys. Rev. Lett.}\ }\textbf {\bibinfo
  {volume} {81}},\ \bibinfo {pages} {2514} (\bibinfo {year}
  {1998})}\BibitemShut {NoStop}%
\bibitem [{\citenamefont {Kozik}\ \emph {et~al.}(2010)\citenamefont {Kozik}
  \emph {et~al.}}]{Kozik10}%
  \BibitemOpen
  \bibfield  {author} {\bibinfo {author} {\bibfnamefont {E.}~\bibnamefont
  {Kozik}} \emph {et~al.},\ }\href {\doibase 10.1209/0295-5075/90/10004}
  {\bibfield  {journal} {\bibinfo  {journal} {Europhys. Lett.}\ }\textbf
  {\bibinfo {volume} {90}},\ \bibinfo {pages} {10004} (\bibinfo {year}
  {2010})}\BibitemShut {NoStop}%
\bibitem [{\citenamefont {Rubtsov}\ \emph {et~al.}(2005)\citenamefont
  {Rubtsov}, \citenamefont {Savkin},\ and\ \citenamefont
  {Lichtenstein}}]{Rubtsov05}%
  \BibitemOpen
  \bibfield  {author} {\bibinfo {author} {\bibfnamefont {A.~N.}\ \bibnamefont
  {Rubtsov}}, \bibinfo {author} {\bibfnamefont {V.~V.}\ \bibnamefont {Savkin}},
  \ and\ \bibinfo {author} {\bibfnamefont {A.~I.}\ \bibnamefont
  {Lichtenstein}},\ }\href {\doibase 10.1103/PhysRevB.72.035122} {\bibfield
  {journal} {\bibinfo  {journal} {Phys. Rev. B}\ }\textbf {\bibinfo {volume}
  {72}},\ \bibinfo {eid} {035122} (\bibinfo {year} {2005})}\BibitemShut
  {NoStop}%
\bibitem [{\citenamefont {Werner}\ \emph {et~al.}(2006)\citenamefont {Werner}
  \emph {et~al.}}]{Werner06}%
  \BibitemOpen
  \bibfield  {author} {\bibinfo {author} {\bibfnamefont {P.}~\bibnamefont
  {Werner}} \emph {et~al.},\ }\href {\doibase 10.1103/PhysRevLett.97.076405}
  {\bibfield  {journal} {\bibinfo  {journal} {Phys. Rev. Lett.}\ }\textbf
  {\bibinfo {volume} {97}},\ \bibinfo {eid} {076405} (\bibinfo {year}
  {2006})}\BibitemShut {NoStop}%
\bibitem [{\citenamefont {Gull}\ \emph {et~al.}(2008)\citenamefont {Gull} \emph
  {et~al.}}]{Gull08_ctaux}%
  \BibitemOpen
  \bibfield  {author} {\bibinfo {author} {\bibfnamefont {E.}~\bibnamefont
  {Gull}} \emph {et~al.},\ }\href {\doibase 10.1209/0295-5075/82/57003}
  {\bibfield  {journal} {\bibinfo  {journal} {Europhys. Lett.}\ }\textbf
  {\bibinfo {volume} {82}},\ \bibinfo {pages} {57003} (\bibinfo {year}
  {2008})}\BibitemShut {NoStop}%
\bibitem [{\citenamefont {Gull}\ \emph {et~al.}(2011)\citenamefont {Gull} \emph
  {et~al.}}]{GullRMP}%
  \BibitemOpen
  \bibfield  {author} {\bibinfo {author} {\bibfnamefont {E.}~\bibnamefont
  {Gull}} \emph {et~al.},\ }\href {\doibase 10.1103/RevModPhys.83.349}
  {\bibfield  {journal} {\bibinfo  {journal} {Rev. Mod. Phys.}\ }\textbf
  {\bibinfo {volume} {83}},\ \bibinfo {pages} {349} (\bibinfo {year}
  {2011})}\BibitemShut {NoStop}%
\bibitem [{\citenamefont {Hanson}\ \emph {et~al.}(2007)\citenamefont {Hanson}
  \emph {et~al.}}]{Hanson07}%
  \BibitemOpen
  \bibfield  {author} {\bibinfo {author} {\bibfnamefont {R.}~\bibnamefont
  {Hanson}} \emph {et~al.},\ }\href {\doibase 10.1103/RevModPhys.79.1217}
  {\bibfield  {journal} {\bibinfo  {journal} {Rev. Mod. Phys.}\ }\textbf
  {\bibinfo {volume} {79}},\ \bibinfo {eid} {1217} (\bibinfo {year}
  {2007})}\BibitemShut {NoStop}%
\bibitem [{\citenamefont {Brako}\ and\ \citenamefont {Newns}(1981)}]{Brako81}%
  \BibitemOpen
  \bibfield  {author} {\bibinfo {author} {\bibfnamefont {R.}~\bibnamefont
  {Brako}}\ and\ \bibinfo {author} {\bibfnamefont {D.~M.}\ \bibnamefont
  {Newns}},\ }\href {\doibase 10.1088/0022-3719/14/21/023} {\bibfield
  {journal} {\bibinfo  {journal} {J. Phys. C}\ }\textbf {\bibinfo {volume}
  {14}},\ \bibinfo {pages} {3065} (\bibinfo {year} {1981})}\BibitemShut
  {NoStop}%
\bibitem [{\citenamefont {Anderson}(1961)}]{Anderson61}%
  \BibitemOpen
  \bibfield  {author} {\bibinfo {author} {\bibfnamefont {P.~W.}\ \bibnamefont
  {Anderson}},\ }\href {\doibase 10.1103/PhysRev.124.41} {\bibfield  {journal}
  {\bibinfo  {journal} {Phys. Rev.}\ }\textbf {\bibinfo {volume} {124}},\
  \bibinfo {pages} {41} (\bibinfo {year} {1961})}\BibitemShut {NoStop}%
\bibitem [{\citenamefont {Georges}\ \emph {et~al.}(1996)\citenamefont {Georges}
  \emph {et~al.}}]{Georges96}%
  \BibitemOpen
  \bibfield  {author} {\bibinfo {author} {\bibfnamefont {A.}~\bibnamefont
  {Georges}} \emph {et~al.},\ }\href {\doibase 10.1103/RevModPhys.68.13}
  {\bibfield  {journal} {\bibinfo  {journal} {Rev. Mod. Phys.}\ }\textbf
  {\bibinfo {volume} {68}},\ \bibinfo {pages} {13} (\bibinfo {year}
  {1996})}\BibitemShut {NoStop}%
\bibitem [{\citenamefont {Maier}\ \emph {et~al.}(2005)\citenamefont {Maier}
  \emph {et~al.}}]{Maier05}%
  \BibitemOpen
  \bibfield  {author} {\bibinfo {author} {\bibfnamefont {T.}~\bibnamefont
  {Maier}} \emph {et~al.},\ }\href {\doibase 10.1103/RevModPhys.77.1027}
  {\bibfield  {journal} {\bibinfo  {journal} {Rev. Mod. Phys.}\ }\textbf
  {\bibinfo {volume} {77}},\ \bibinfo {eid} {1027} (\bibinfo {year}
  {2005})}\BibitemShut {NoStop}%
\bibitem [{\citenamefont {Gull}\ \emph {et~al.}(2007)\citenamefont {Gull},
  \citenamefont {Werner}, \citenamefont {Millis},\ and\ \citenamefont
  {Troyer}}]{Gull07}%
  \BibitemOpen
  \bibfield  {author} {\bibinfo {author} {\bibfnamefont {E.}~\bibnamefont
  {Gull}}, \bibinfo {author} {\bibfnamefont {P.}~\bibnamefont {Werner}},
  \bibinfo {author} {\bibfnamefont {A.}~\bibnamefont {Millis}}, \ and\ \bibinfo
  {author} {\bibfnamefont {M.}~\bibnamefont {Troyer}},\ }\href {\doibase
  10.1103/PhysRevB.76.235123} {\bibfield  {journal} {\bibinfo  {journal} {Phys.
  Rev. B}\ }\textbf {\bibinfo {volume} {76}},\ \bibinfo {eid} {235123}
  (\bibinfo {year} {2007})}\BibitemShut {NoStop}%
\bibitem [{\citenamefont {M\"uhlbacher}\ and\ \citenamefont
  {Rabani}(2008)}]{Muehlbacher08}%
  \BibitemOpen
  \bibfield  {author} {\bibinfo {author} {\bibfnamefont {L.}~\bibnamefont
  {M\"uhlbacher}}\ and\ \bibinfo {author} {\bibfnamefont {E.}~\bibnamefont
  {Rabani}},\ }\href {\doibase 10.1103/PhysRevLett.100.176403} {\bibfield
  {journal} {\bibinfo  {journal} {Phys. Rev. Lett.}\ }\textbf {\bibinfo
  {volume} {100}},\ \bibinfo {pages} {176403} (\bibinfo {year}
  {2008})}\BibitemShut {NoStop}%
\bibitem [{\citenamefont {Schir\'o}\ and\ \citenamefont
  {Fabrizio}(2009)}]{Schiro09}%
  \BibitemOpen
  \bibfield  {author} {\bibinfo {author} {\bibfnamefont {M.}~\bibnamefont
  {Schir\'o}}\ and\ \bibinfo {author} {\bibfnamefont {M.}~\bibnamefont
  {Fabrizio}},\ }\href {\doibase 10.1103/PhysRevB.79.153302} {\bibfield
  {journal} {\bibinfo  {journal} {Phys. Rev. B}\ }\textbf {\bibinfo {volume}
  {79}},\ \bibinfo {pages} {153302} (\bibinfo {year} {2009})}\BibitemShut
  {NoStop}%
\bibitem [{\citenamefont {Werner}\ \emph {et~al.}(2009)\citenamefont {Werner},
  \citenamefont {Oka},\ and\ \citenamefont {Millis}}]{Werner09}%
  \BibitemOpen
  \bibfield  {author} {\bibinfo {author} {\bibfnamefont {P.}~\bibnamefont
  {Werner}}, \bibinfo {author} {\bibfnamefont {T.}~\bibnamefont {Oka}}, \ and\
  \bibinfo {author} {\bibfnamefont {A.~J.}\ \bibnamefont {Millis}},\ }\href
  {\doibase 10.1103/PhysRevB.79.035320} {\bibfield  {journal} {\bibinfo
  {journal} {Phys. Rev. B}\ }\textbf {\bibinfo {volume} {79}},\ \bibinfo {eid}
  {035320} (\bibinfo {year} {2009})}\BibitemShut {NoStop}%
\bibitem [{\citenamefont {Werner}\ \emph {et~al.}(2010)\citenamefont {Werner}
  \emph {et~al.}}]{Werner10}%
  \BibitemOpen
  \bibfield  {author} {\bibinfo {author} {\bibfnamefont {P.}~\bibnamefont
  {Werner}} \emph {et~al.},\ }\href {\doibase 10.1103/PhysRevB.81.035108}
  {\bibfield  {journal} {\bibinfo  {journal} {Phys. Rev. B}\ }\textbf {\bibinfo
  {volume} {81}},\ \bibinfo {pages} {035108} (\bibinfo {year}
  {2010})}\BibitemShut {NoStop}%
\bibitem [{\citenamefont {Gull}\ \emph {et~al.}(2010)\citenamefont {Gull},
  \citenamefont {Reichman},\ and\ \citenamefont {Millis}}]{Gull10_bold}%
  \BibitemOpen
  \bibfield  {author} {\bibinfo {author} {\bibfnamefont {E.}~\bibnamefont
  {Gull}}, \bibinfo {author} {\bibfnamefont {D.~R.}\ \bibnamefont {Reichman}},
  \ and\ \bibinfo {author} {\bibfnamefont {A.~J.}\ \bibnamefont {Millis}},\
  }\href {\doibase 10.1103/PhysRevB.82.075109} {\bibfield  {journal} {\bibinfo
  {journal} {Phys. Rev. B}\ }\textbf {\bibinfo {volume} {82}},\ \bibinfo
  {pages} {075109} (\bibinfo {year} {2010})}\BibitemShut {NoStop}%
\bibitem [{\citenamefont {Prokof'ev}\ and\ \citenamefont
  {Svistunov}(2007)}]{Prokofev07}%
  \BibitemOpen
  \bibfield  {author} {\bibinfo {author} {\bibfnamefont {N.}~\bibnamefont
  {Prokof'ev}}\ and\ \bibinfo {author} {\bibfnamefont {B.}~\bibnamefont
  {Svistunov}},\ }\href {\doibase 10.1103/PhysRevLett.99.250201} {\bibfield
  {journal} {\bibinfo  {journal} {Phys. Rev. Lett.}\ }\textbf {\bibinfo
  {volume} {99}},\ \bibinfo {eid} {250201} (\bibinfo {year}
  {2007})}\BibitemShut {NoStop}%
\bibitem [{\citenamefont {Prokof'ev}\ and\ \citenamefont
  {Svistunov}(2008)}]{Prokofev08B}%
  \BibitemOpen
  \bibfield  {author} {\bibinfo {author} {\bibfnamefont {N.~V.}\ \bibnamefont
  {Prokof'ev}}\ and\ \bibinfo {author} {\bibfnamefont {B.~V.}\ \bibnamefont
  {Svistunov}},\ }\href {\doibase 10.1103/PhysRevB.77.125101} {\bibfield
  {journal} {\bibinfo  {journal} {Phys. Rev. B}\ }\textbf {\bibinfo {volume}
  {77}},\ \bibinfo {eid} {125101} (\bibinfo {year} {2008})}\BibitemShut
  {NoStop}%
\bibitem [{\citenamefont {Keiter}\ and\ \citenamefont
  {Kimball}(1971)}]{Keiter71}%
  \BibitemOpen
  \bibfield  {author} {\bibinfo {author} {\bibfnamefont {H.}~\bibnamefont
  {Keiter}}\ and\ \bibinfo {author} {\bibfnamefont {J.}~\bibnamefont
  {Kimball}},\ }\href@noop {} {\bibfield  {journal} {\bibinfo  {journal} {Int.
  J. Magn.}\ }\textbf {\bibinfo {volume} {1}},\ \bibinfo {pages} {233}
  (\bibinfo {year} {1971})}\BibitemShut {NoStop}%
\bibitem [{\citenamefont {Bickers}(1987)}]{Bickers87}%
  \BibitemOpen
  \bibfield  {author} {\bibinfo {author} {\bibfnamefont {N.~E.}\ \bibnamefont
  {Bickers}},\ }\href {\doibase 10.1103/RevModPhys.59.845} {\bibfield
  {journal} {\bibinfo  {journal} {Rev. Mod. Phys.}\ }\textbf {\bibinfo {volume}
  {59}},\ \bibinfo {pages} {845} (\bibinfo {year} {1987})}\BibitemShut
  {NoStop}%
\bibitem [{\citenamefont {Bauer}\ \emph {et~al.}(2011)\citenamefont {Bauer}
  \emph {et~al.}}]{ALPS2}%
  \BibitemOpen
  \bibfield  {author} {\bibinfo {author} {\bibfnamefont {B.}~\bibnamefont
  {Bauer}} \emph {et~al.},\ }\href
  {http://stacks.iop.org/1742-5468/2011/i=05/a=P05001} {\bibfield  {journal}
  {\bibinfo  {journal} {JSTAT}\ }\textbf {\bibinfo {volume} {2011}},\ \bibinfo
  {pages} {P05001} (\bibinfo {year} {2011})}\BibitemShut {NoStop}%
\end{thebibliography}%

\end{document}